\documentclass[twocolumn]{IEEEtran2e} %!PN
\usepackage{epsfig}
\usepackage{rotating}

\begin{document}

\title{Construction and Test of the Precision Drift Chambers for the ATLAS Muon Spectrometer
}

\author{
F.~Bauer, W.~Blum, U.~Bratzler, H.~Dietl, S.~Kotov, H.~Kroha,
Th.~Lagouri, A.~Manz, A.~Ostapchuk, R.~Richter, S.~Schael\\
{\small Max-Planck-Institut f\"ur Physik, F\"ohringer Ring 6, D-80805 Munich, Germany}\\
S.~Chouridou, M.~Deile, O.~Kortner, A.~Staude,
R.~Str\"ohmer, T.~Trefzger\\
{\small Ludwig-Maximilians-Universit\"at, Schellingstra\ss e 4, D-80799 Munich, Germany.}
\thanks{Corresponding author: H.~Kroha, Max-Planck-Institut f\"ur Physik,
F\"ohringer Ring 6, D-80805 Munich, Germany (e-mail: kroha@mppmu.mpg.de).}
\thanks{Permanent address of F.~Bauer: CEA Saclay, DSM, DAPNIA, F-91191 Gif-sur-Yvette Cedex,
France.}
\thanks{Permanent address of S.~Kotov: Joint Institute for Nuclear Research, Dubna,
141980 Moscow Region, Russia.}
}
\maketitle

\begin{abstract}
The Monitored Drift Tube (MDT) chambers for the muon spectrometer
of the ATLAS detector at the Large Hadron Collider (LHC) consist of
3--4 layers of pressurised drift tubes on either side of
a space frame carrying an optical deformation monitoring system.
The chambers have to provide a track position resolution of $40~\mu$m
with a single-tube resolution of at least $80~\mu$m and
a sense wire positioning accuracy of $20~\mu$m (rms).
The feasibility was demonstrated with the full-scale prototype
of one of the largest MDT chambers with 432 drift tubes of 3.8~m length.
For the ATLAS muon spectrometer, 88 chambers of this type have to be built.
The first chamber has been completed with a wire positioning accuracy
of $14~\mu$m (rms).
\end{abstract}

\begin{keywords}
Drift tubes, drift chambers, muon spectrometer, ATLAS detector
\end{keywords}

\section{Introduction}
The muon spectrometer of the ATLAS experiment~\cite{TDR} will be
operated in the toroidal magnetic field of a superconducting air-core
magnet system with 3--6~Tm bending power.It is designed to provide
stand-alone muon momentum resolution of $\Delta p_T/p_T = 2-10~\%$
for transverse momenta between 6~GeV and 1~TeV over a pseudo-rapidity
range of $|\eta |\le 2.7$.  
This requires very accurate track sagitta measurement with three layers of
muon chambers and high-precision optical alignment monitoring systems.
Precision drift chambers, the Monitored Drift Tube (MDT) chambers, 
have been developed to provide a track position resolution of $40~\mu$m 
over an active area of 5500~m${}^2$.

\begin{figure}[h]
\vspace{3mm}

\begin{center}
%\mbox{\epsfig{figure=fmmdtch.eps,angle=-90,width=\linewidth}}
\includegraphics[width=\linewidth]{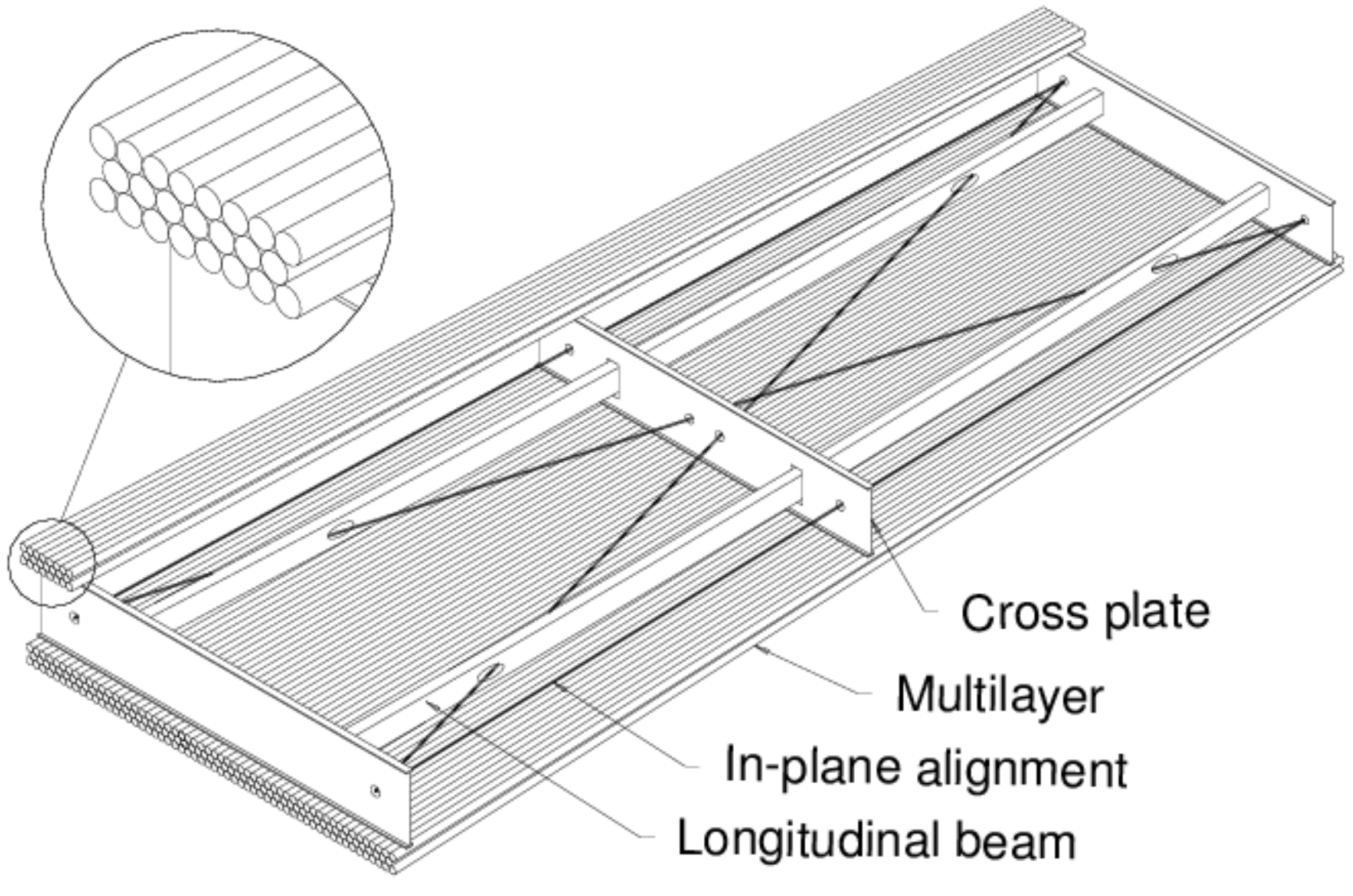}
\end{center}
\caption{Monitored Drift Tube (MDT) chamber for the ATLAS muon spectrometer.
\label{MDT}
}
\end{figure}
\begin{figure}[h!]
\vspace{-2mm}

\begin{center}
%\mbox{\epsfig{figure=/.at/WWW/ftp/outgoing/mdt/Endplugs/Documentation/figure01.eps,
%angle=90,width=0.98\linewidth}}
%\mbox{\epsfig{figure=/.at/WWW/ftp/outgoing/mdt/Endplugs/Endplug99/wire-pos.eps,
%angle=95,width=1.00\linewidth}}
\includegraphics[angle=90,width=0.98\linewidth]{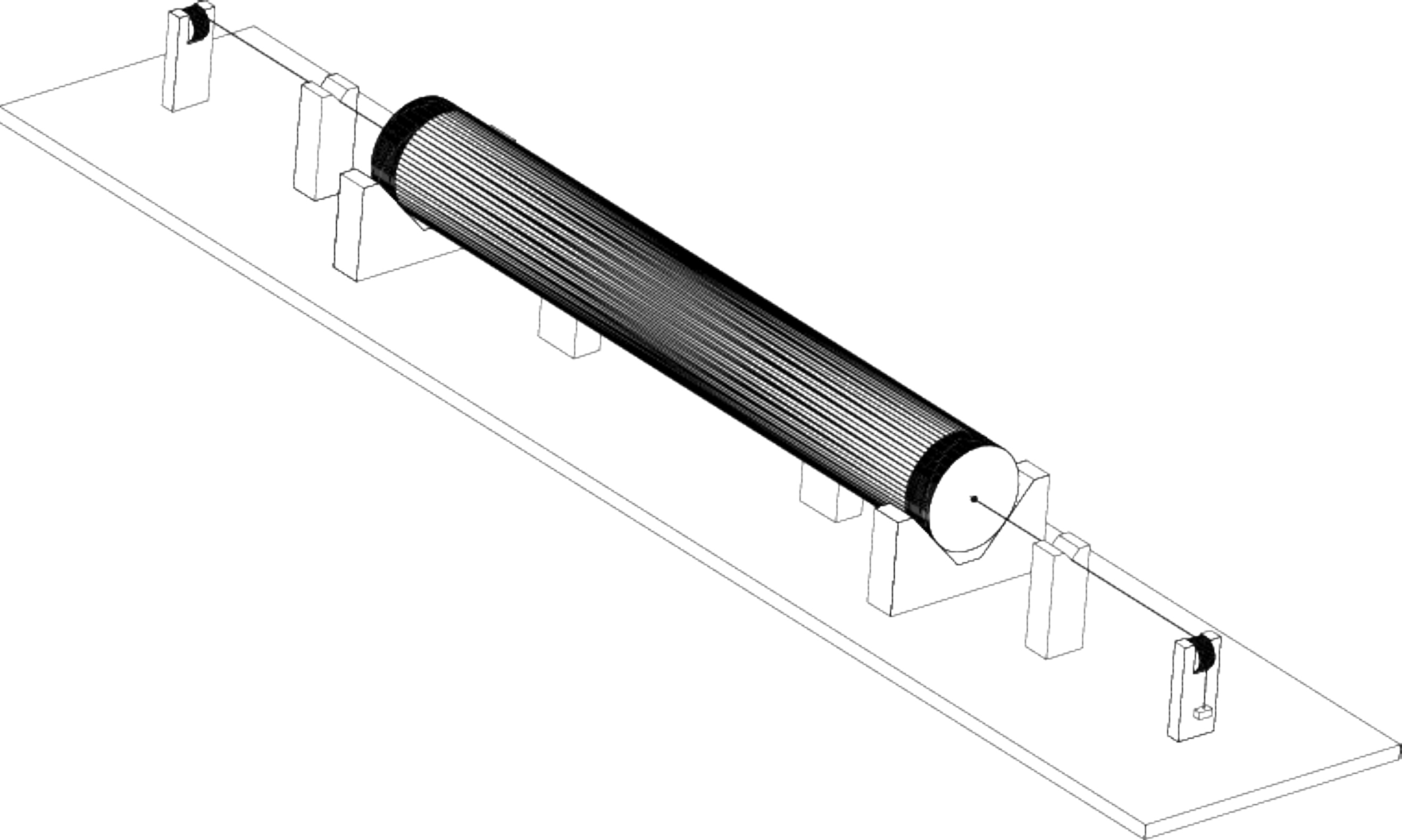}
\end{center}

\vspace{2mm}
\caption{Wire positioning method for the drift tubes of the prototype chamber.
\label{glue}
}
%\end{figure}
%
%
%\begin{figure}[h]
\vspace{4mm}

\hspace{-5mm}
%\mbox{\epsfig{figure=/.at/WWW/ftp/outgoing/mdt/Endplugs/Documentation/figure04.eps,
%angle=-90,width=1.07\linewidth}}
%\mbox{\epsfig{figure=/.at/WWW/ftp/outgoing/mdt/Endplugs/Documentation/figure07.eps,
%angle=90,width=1.00\linewidth}}
%\mbox{\epsfig{figure=/.at/WWW/ftp/outgoing/mdt/Endplugs/Documentation/ep_glue_cut.eps,
%angle=90,width=1.10\linewidth}}
\includegraphics[angle=90,width=1.10\linewidth]{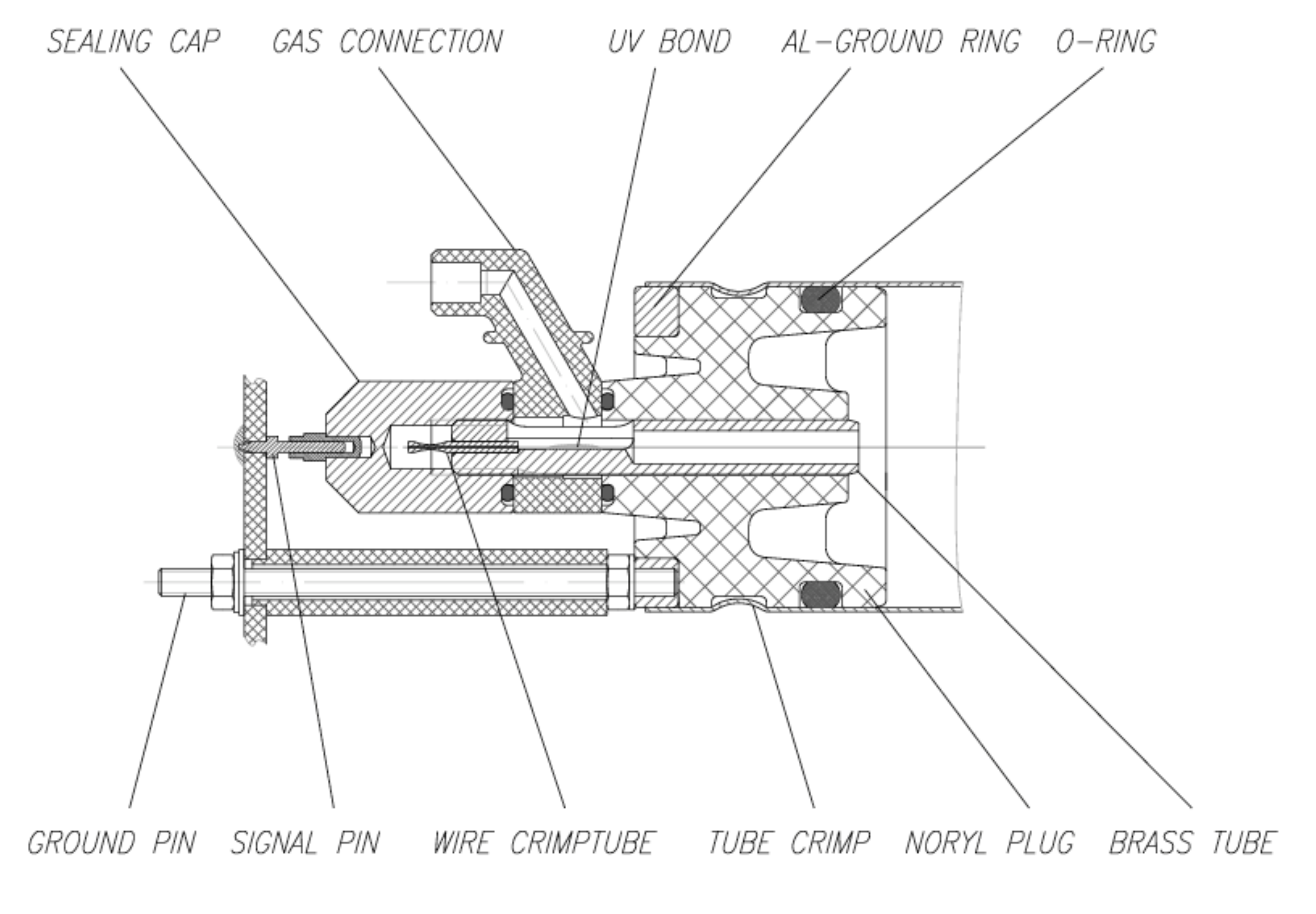}
\caption{Endplug for the glueing technique for precise wire location. The wire is glued
in the groove of the central brass tube.
\label{endplug1}
}
\end{figure}
\begin{figure}[h!]
\begin{center}
%\mbox{\epsfig{figure=/home/iwsatlas1/kroha/tex/atlas/qaqc/mdt_wire_pos_2d_bwt.eps,
%width=\linewidth}}
\includegraphics[width=1.08\linewidth]{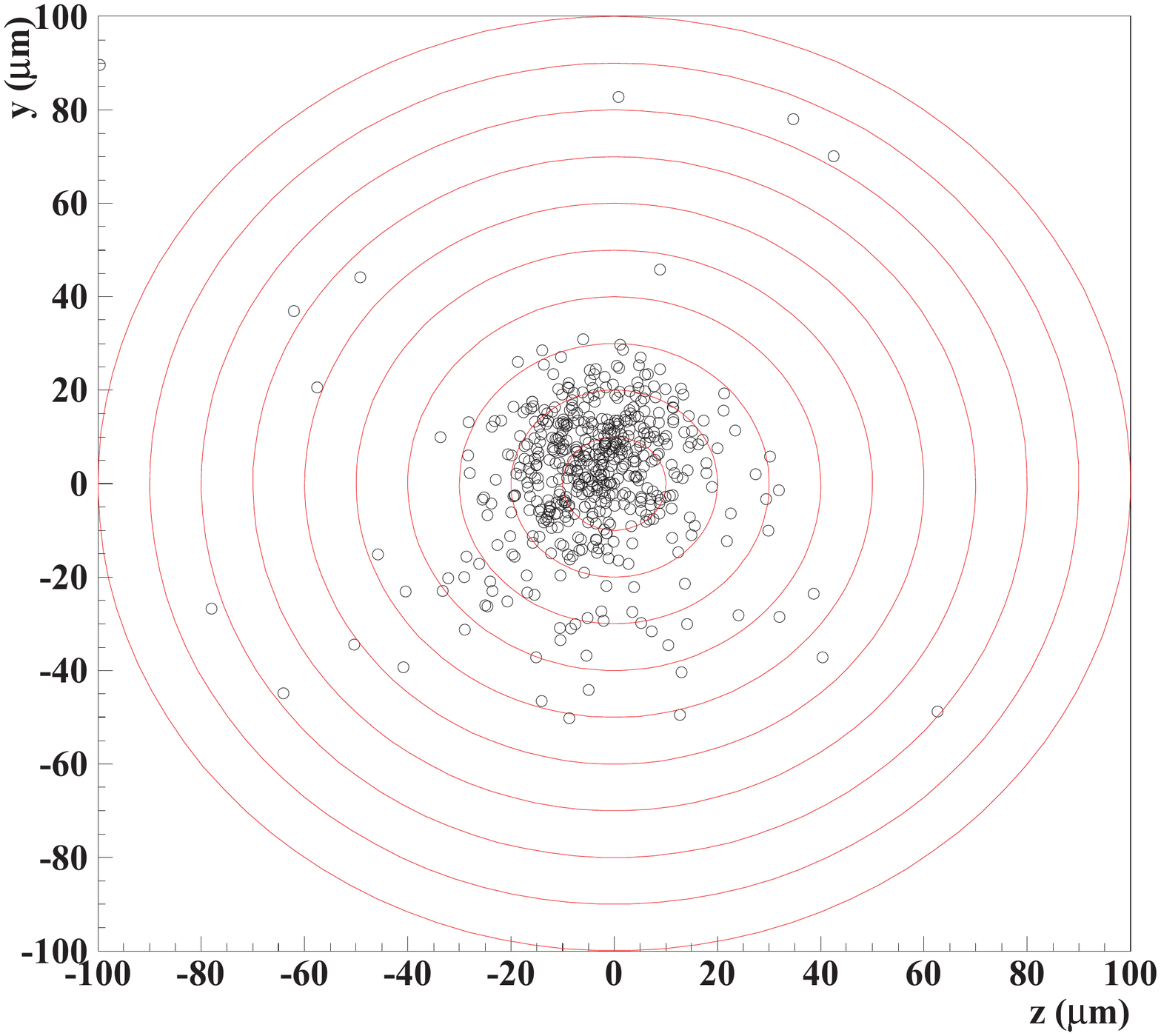}
\end{center}
\caption{Wire positioning accuracy with the glueing technique.
\label{wp1_2d}
}
%\end{figure}
%
%
%\begin{figure}[h]
\vspace{-2mm}

\hspace{-5mm}
\begin{center}
%\mbox{\epsfig{figure=/.at/WWW/ftp/outgoing/mdt/Endplugs/Endplug99/endplug99_new.eps,
%angle=-90,width=1.08\linewidth}}
%\mbox{\epsfig{figure=/home/pcatlas17/kroha/eps/ep99_3d_bw.eps,
%width=1.05\linewidth}}
%\mbox{\epsfig{figure=/.at/WWW/ftp/outgoing/mdt/Endplugs/Endplug99/ep99_cut_2.eps,
%angle=90,width=1.16\linewidth}}
\includegraphics[angle=90,width=1.18\linewidth]{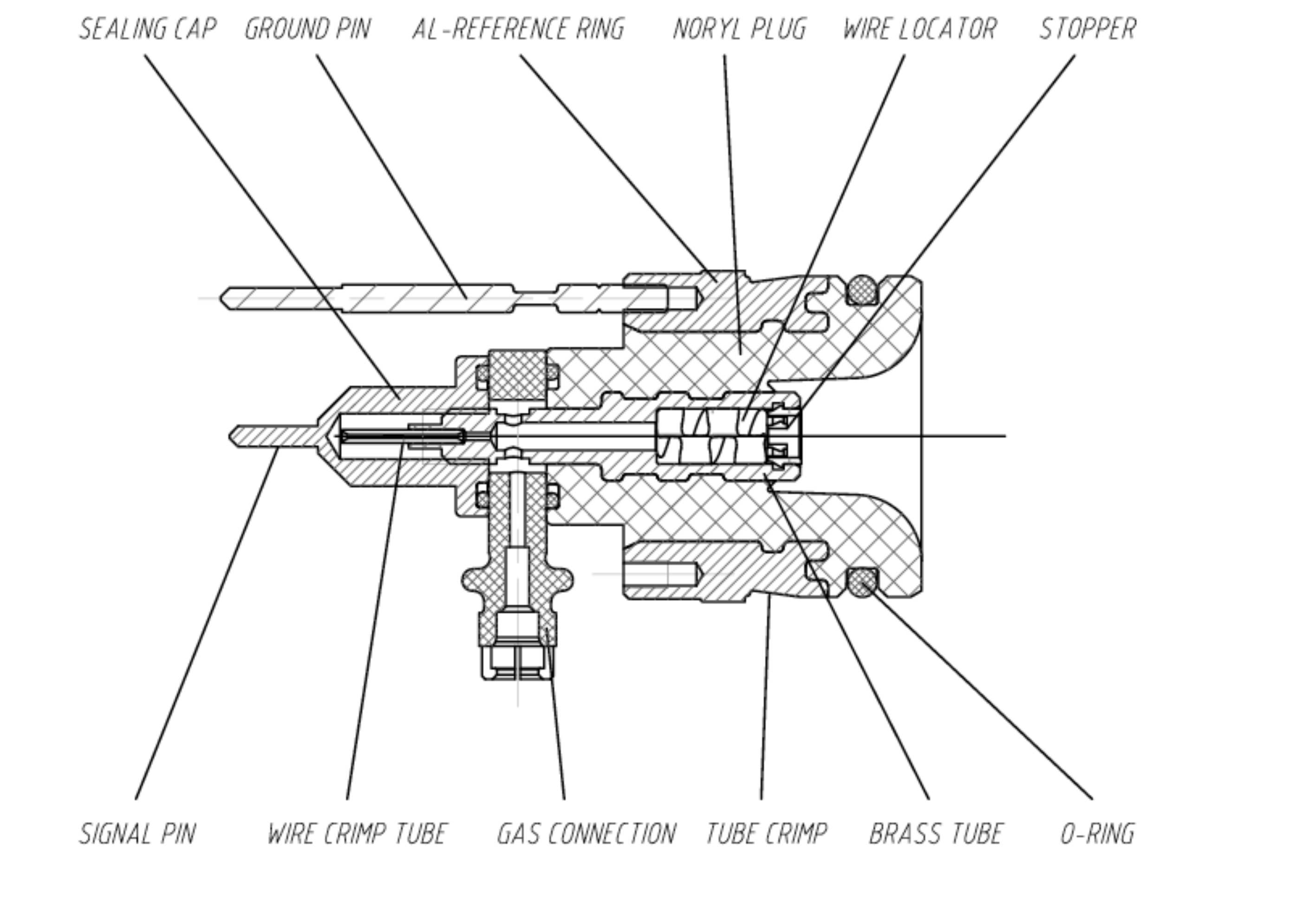}
\end{center}
\vspace{-4mm}
\caption{Endplug with precise mechanical wire location with respect
to the aluminum reference ring.
\label{endplug2}
}
\end{figure}
\begin{figure}[h]
\vspace{-4mm}

%\mbox{\epsfig{figure=/home/iwsatlas1/kroha/tex/atlas/qaqc/bos1_tubes_xray_dubna_b1d_2_bwt.eps,
%width=1.10\linewidth}}
\hspace{-15mm}
\includegraphics[width=1.25\linewidth]{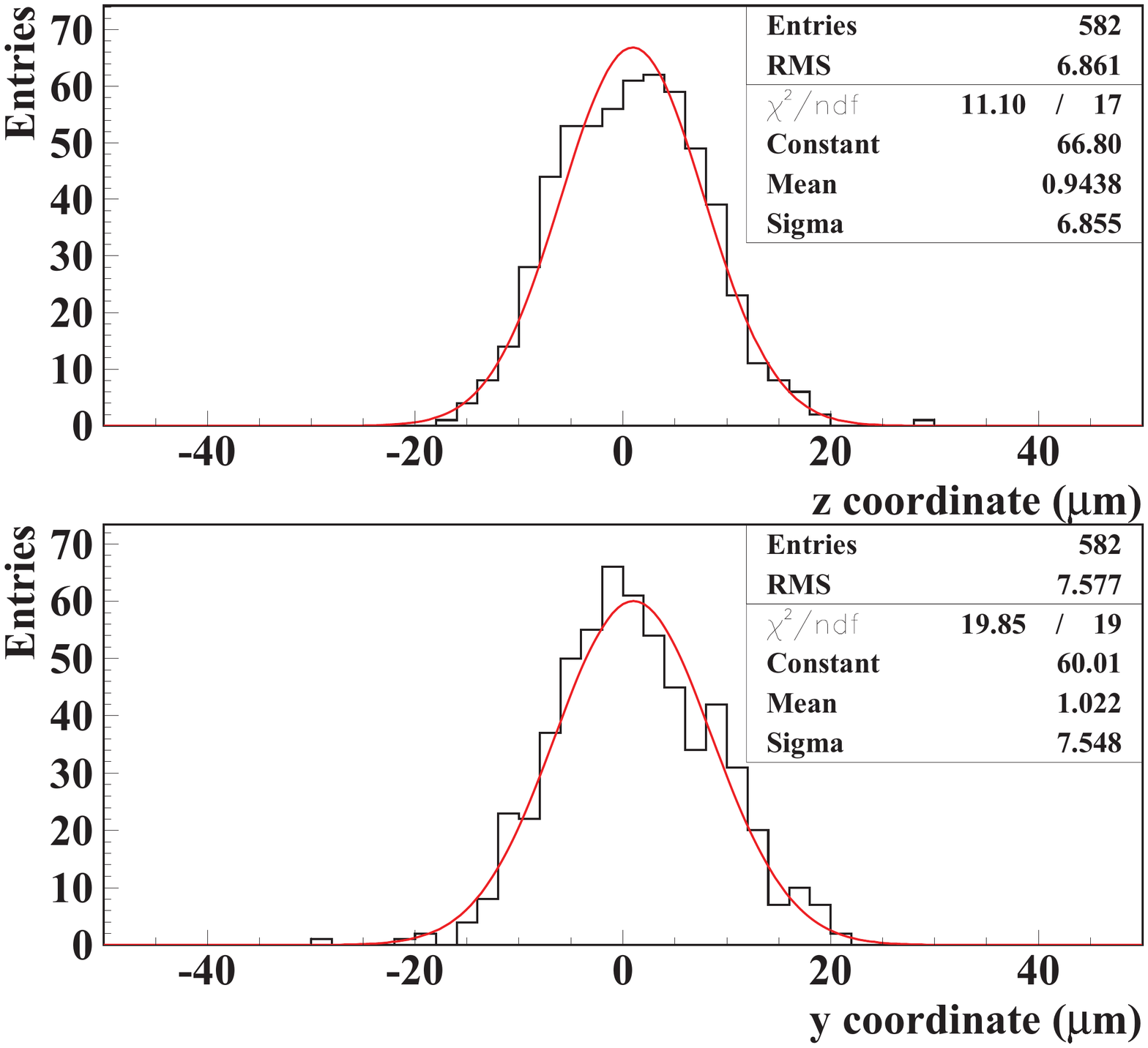}

\vspace{-5mm}
\caption{Wire positioning accuracy in individual drift tubes 
with the precise endplugs used for chamber serial production in the z coordinate parallel
to the tube layers (top) and in the y coordinate perpendicular to the tube layers (bottom).
\vspace{-7mm}
\label{wp2_1d}
}

%\begin{large}
%\vspace{-90mm}\hspace{10mm} {\bf (a)}

%\vspace{36mm}\hspace{10mm} {\bf (b)}

%\vspace{49mm}
%\end{large}
\end{figure}
\begin{table}[h]

\caption{Geometrical chamber parameters (see text) from the fit to the X-ray measurements
(with rms errors) in comparison with the design values (with absolute tolerances)
\label{fit}}

\begin{center}
\begin{tabular}{lcc}

\hline
\multicolumn{3}{c} {\bf BOS~98 High Voltage End} \\
\hline

Parameter & X-ray fit & Design value \\
\hline

$y$-pitch $[\mu$m$]$ & $\hphantom{.2}26058\pm 0.5\hphantom{01}$ & $\hphantom{.2}26054\pm 5\hphantom{001.}$ \\

$z$-pitch $[\mu$m$]$ & $30036.2\pm 0.5\hphantom{01}$ & $\hphantom{.2}30036\pm 0.5\hphantom{01}$ \\
\hline

$\Delta y$ $[$mm$]$  & $347.041\pm 0.011$ & $347.040\pm 0.010$ \\

$\Delta z$ $[\mu$m$]$  & $\hphantom{347}-8\pm 11\hphantom{0.0}$ & $\hphantom{346.87}0\pm 10\hphantom{0.0}$ \\

\hline
\multicolumn{3}{c} {\bf BOS~98 Readout End} \\
\hline

Parameter & X-ray fit & Design value \\
\hline

$y$-pitch $[\mu$m$]$ & $\hphantom{.2}26060\pm 1.3\hphantom{01}$ & $\hphantom{.2}26054\pm 5\hphantom{001.}$ \\

$z$-pitch $[\mu$m$]$ & $30036.2\pm 0.5\hphantom{01}$ & $\hphantom{.2}30036\pm 0.5\hphantom{01}$ \\
\hline

$\Delta y$ $[$mm$]$  & $347.078\pm 0.012$ & $347.072\pm 0.010$ \\

$\Delta z$ $[\mu$m$]$  & $\hphantom{346.87}3\pm 16\hphantom{0.0}$ & $\hphantom{346.87}0\pm 10\hphantom{0.0}$ \\
\hline\hline
& & \\ [-2mm]
\hline
\multicolumn{3}{c} {\bf BOS-0 High Voltage End} \\
\hline

Parameter & X-ray fit & Design value \\
\hline

$y$-pitch $[\mu$m$]$ & $\hphantom{.8}26039\pm 0.5\hphantom{01}$ & $\hphantom{.2}26039\pm 5\hphantom{001.}$ \\

$z$-pitch $[\mu$m$]$ & $30035.8\pm 0.5\hphantom{01}$ & $\hphantom{.2}30036\pm 0.5\hphantom{01}$ \\
\hline

$\Delta y$ $[$mm$]$  & $346.878\pm 0.005$ & $346.882\pm 0.010$ \\

$\Delta z$ $[\mu$m$]$  & $\hphantom{34}-22\pm 5\hphantom{0.01}$ & $\hphantom{346.87}0\pm 10\hphantom{0.0}$ \\

\hline
\multicolumn{3}{c} {\bf BOS-0 Readout End} \\
\hline

Parameter & X-ray fit & Design value \\
\hline

$y$-pitch $[\mu$m$]$ & $\hphantom{.2}26041\pm 0.5\hphantom{01}$ & $\hphantom{.2}26039\pm 5\hphantom{001.}$ \\

$z$-pitch $[\mu$m$]$ & $30035.8\pm 0.5\hphantom{01}$ & $\hphantom{.2}30036\pm 0.5\hphantom{01}$ \\
\hline

$\Delta y$ $[$mm$]$  & $346.852\pm 0.005$ & $346.882\pm 0.010$ \\

$\Delta z$ $[\mu$m$]$  & $\hphantom{347}-9\pm 5\hphantom{0.01}$ & $\hphantom{346.87}0\pm 10\hphantom{0.0}$ \\
\hline\hline

\end{tabular}
\end{center}
\end{table}
\renewcommand{\tabcolsep}{1.0mm}
\begin{table}[h]

\caption{Widths (rms) of the wire coordinate residual distributions
\label{width}}

\begin{center}
\begin{tabular}{lccc}

\hline
\multicolumn{4}{c}{\bf BOS~98 Chamber} \\
\hline
& High voltage end & Readout end & Center \\
\hline

$y$-coord.~$[\mu$m$]$ & $19.0\pm 0.8$ & $17.0\pm 0.2$ & 15.0 \\

$z$-coord.~$[\mu$m$]$ & $19.3\pm 0.8$ & $18.2\pm 1.3$ & 16.1 \\
\hline

Combined $[\mu$m$]$ & \multicolumn{2}{c}{18.4} & 15.6 \\
\hline\hline
& & \\ [-2mm]
\hline
\multicolumn{4}{c}{\bf BOS-0 Chamber} \\
\hline
& High voltage end & Readout end & Center \\
\hline

$y$-coord.~$[\mu$m$]$ & 14.3 & 15.3 & 13.2 \\

$z$-coord.~$[\mu$m$]$ & 10.5 & 14.1 & \hphantom{1}7.7 \\
\hline

Combined $[\mu$m$]$ & \multicolumn{2}{c}{13.6} & 10.5 \\
\hline\hline

\end{tabular}
\end{center}
\end{table}
\begin{figure}[h]
\vspace{8mm}

\begin{center}
%\mbox{\epsfig{figure=klebe2.eps,width=1.02\linewidth}}
\hspace{-6mm}
\includegraphics[width=1.06\linewidth]{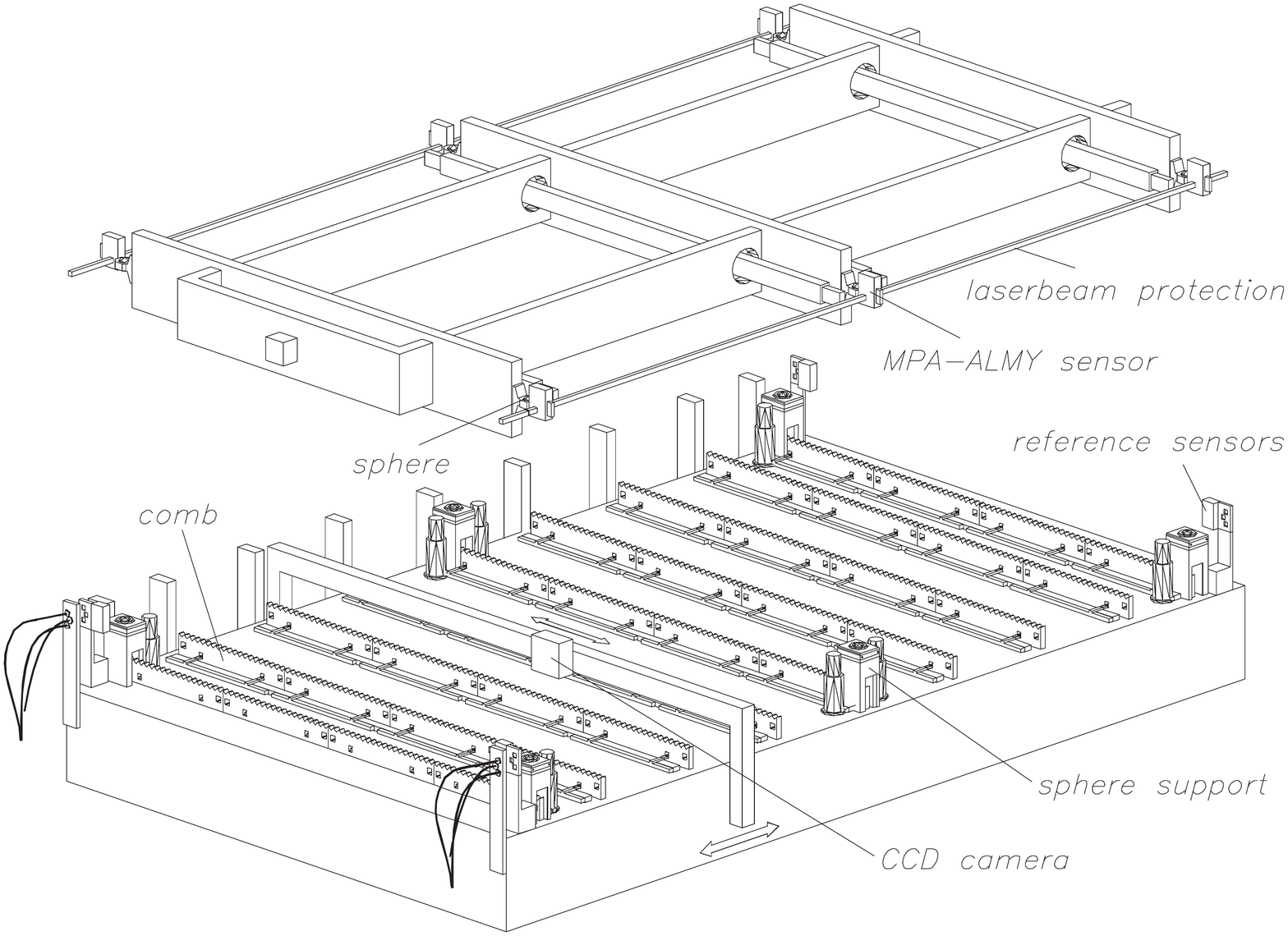}
\end{center}
\caption{Chamber assembly table with precision jigs,
space frame with bars for cross plate sag compensation (see text) and optical monitoring devices.
\label{table1}
}
\end{figure}
%
%\clearpage

%\begin{figure}[h]
%
%\hspace{-7mm}
%\mbox{\epsfig{figure=/.at/wwwatlas/ftp/outgoing/mdt/Module0/assembly/bos-0_sag_crossplate_bw.eps,
%width=1.10\linewidth}}
%\vspace{-8mm}
%\caption{Gravitational sag of the cross plates at the high voltage (HV) and readout (RO) ends and in the middle (MI)
%of the BOS-0 chamber as a function of the number of tube layers glued to the space frame (a) before and
%(b) after the compensation (HV: circles, RO: squares, MI: triangles).
%\label{sag2}
%}
%
%\begin{large}
%\vspace{-94mm}\hspace{18mm} {\bf (a)}
%
%\vspace{36mm}\hspace{18mm} {\bf (b)}
%
%\vspace{47mm}
%\end{large}
%\end{figure}
%
\begin{figure}[h]

\hspace{-7mm}
%\mbox{\epsfig{figure=/.at/wwwatlas/ftp/outgoing/mdt/Module0/assembly/bos-0_sag_crossplate_bwt.eps,
%width=1.18\linewidth}}
\includegraphics[width=1.10\linewidth]{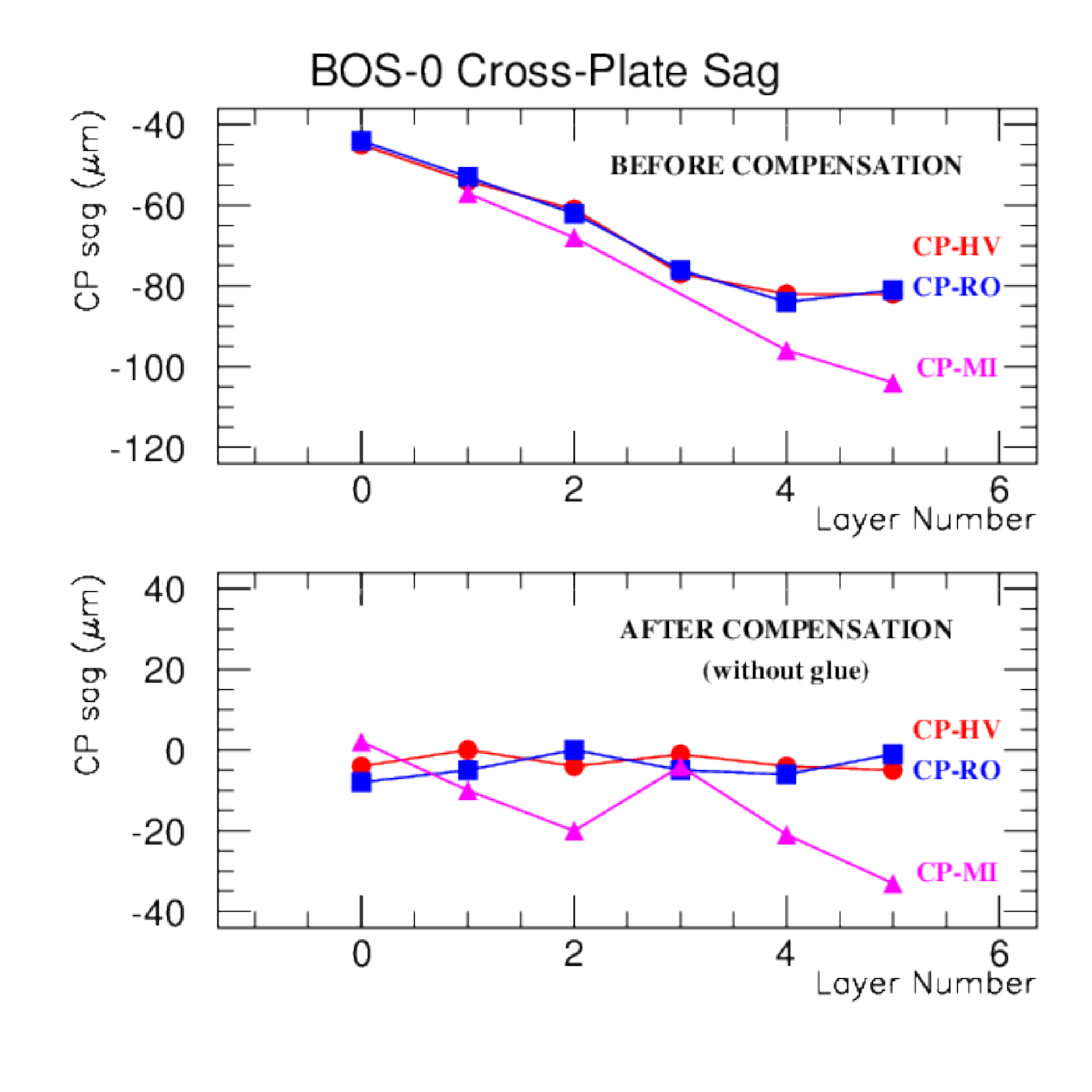}
\vspace{-10mm}
\caption{Gravitational sag of the cross plates at the high voltage (HV) and readout (RO) ends and in the middle (MI)
of the BOS-0 chamber as a function of the number of tube layers glued to the space frame (a) before and
(b) after the compensation (HV: circles, RO: squares, MI: triangles).
\label{sag2}
}

\begin{large}
\vspace{-79mm}\hspace{18mm} {\bf (a)}

\vspace{38mm}\hspace{18mm} {\bf (b)}

\vspace{27mm}
\end{large}
\end{figure}
\begin{figure}[h]
\vspace{-18mm}

\hspace{-2mm}
%\mbox{\epsfig{figure=/home/iwsatlas1/kroha/tex/atlas/fortran/mun98_5_8p_bwt.eps,
%width=1.17\linewidth}}
\includegraphics[width=1.04\linewidth]{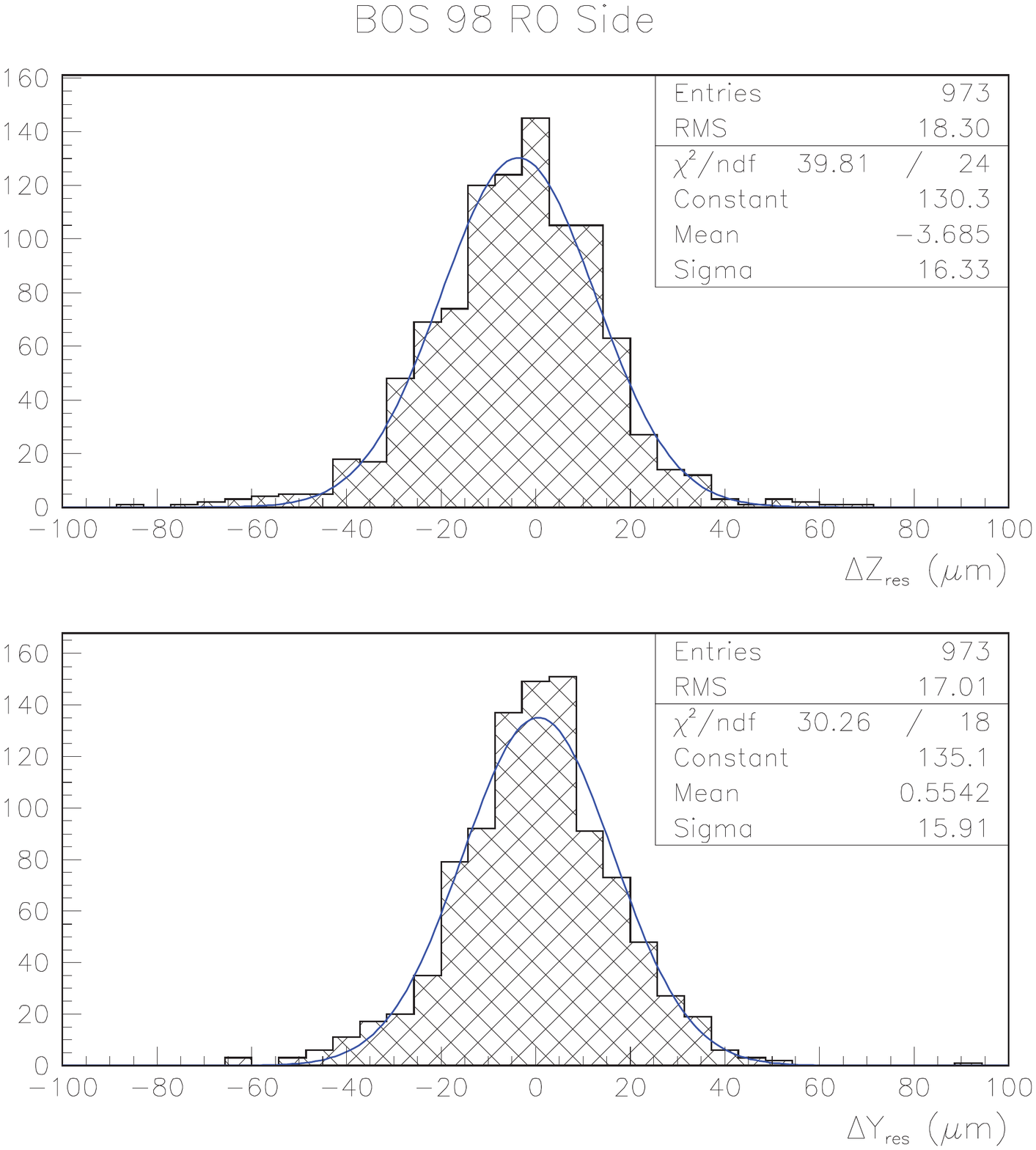}
\vspace{-20mm}

\caption{Distributions of the residuals of the
measured wire positions with respect to the fitted grid
at the readout end of the BOS~98 prototype chamber in the z coordinate parallel to the
tube layers (top) and in the y coordinate perpendicular to the tube layers (bottom).
\label{Xray1}
}

%\begin{large}
%\vspace{-107mm}\hspace{6mm} {\bf (a)}

%\vspace{43mm}\hspace{6mm} {\bf (b)}

%\vspace{65mm}
%\end{large}
\end{figure}
\begin{figure}[h]
%\vspace{-7mm}

\hspace{-2mm}
%\mbox{\epsfig{figure=/home/iwsatlas1/kroha/tex/atlas/fortran/bos_2000_08_p2_r0_25_8p_bwt.eps,
%width=1.24\linewidth}}
\includegraphics[width=1.04\linewidth]{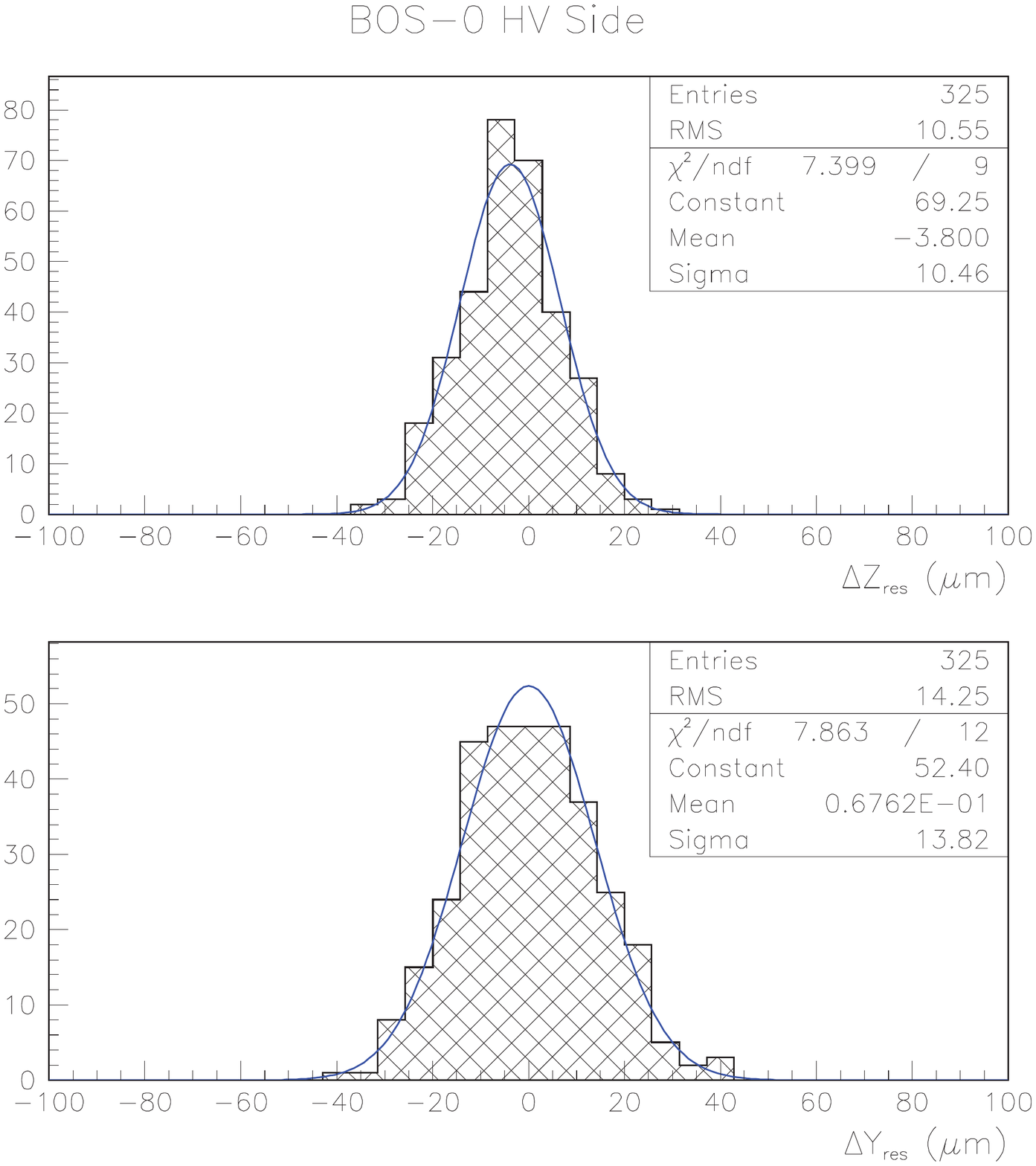}
\vspace{-15mm}

\caption{As Fig.~9 for the high-voltage end of the BOS-0 chamber.
\label{Xray2}
}

%\begin{large}
%\vspace{-98mm}\hspace{6mm} {\bf (a)}

%\vspace{44mm}\hspace{6mm} {\bf (b)}

%\vspace{45mm}
%\end{large}
\end{figure}

\begin{figure}[h]
\vspace{-3mm}

\hspace{-2mm}
%\mbox{\epsfig{figure=/home/iwsatlas1/kroha/tex/atlas/fortran/mun98_5_r2_10_4pp_layer_sct.eps,
%width=1.21\linewidth}}
\includegraphics[width=1.02\linewidth]{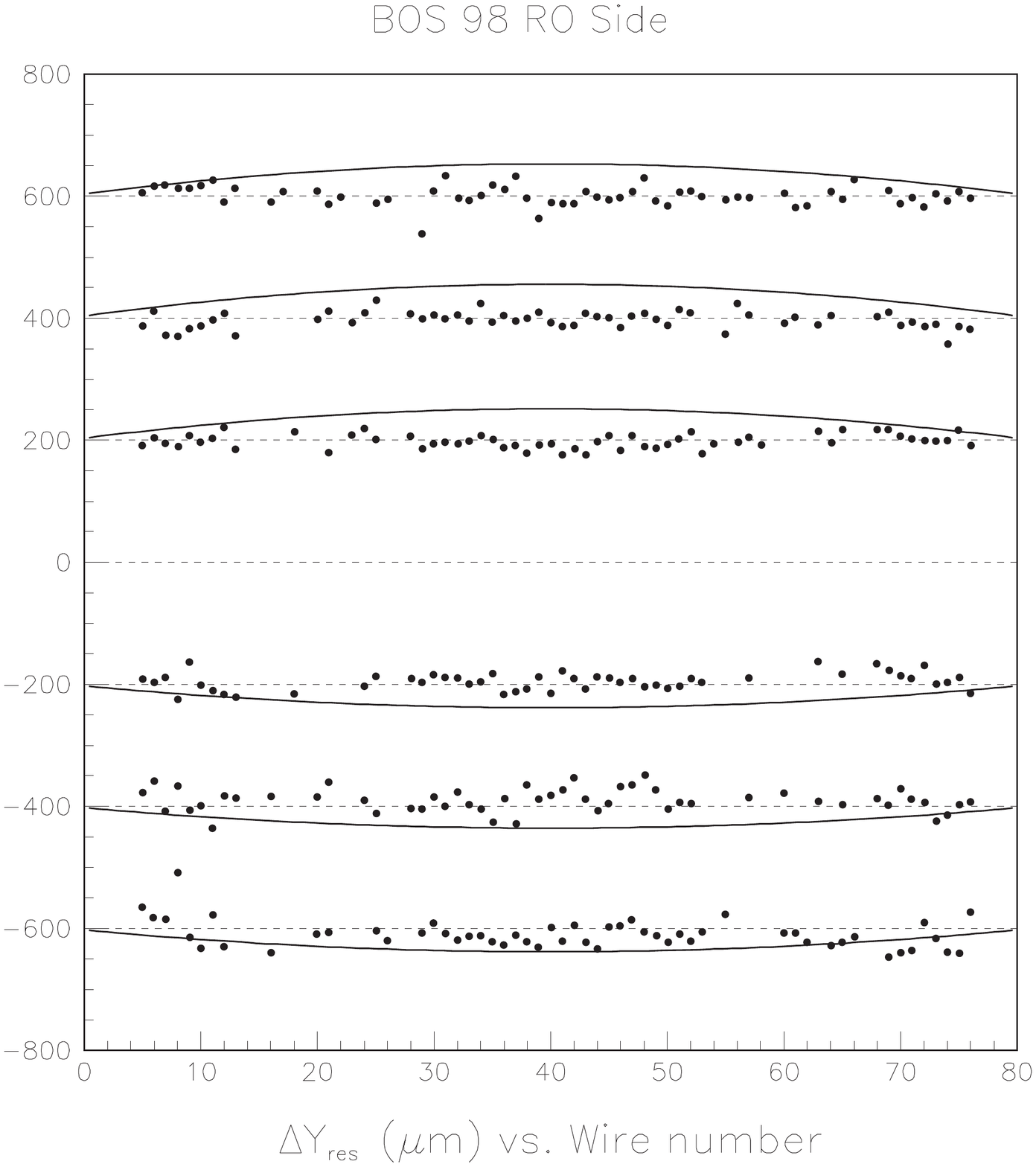}
\vspace{-18mm}

\caption{Residuals $\Delta y_{\mathrm res}$ of the measured $y$ coordinates of the wires
in the 6 tube layers (distances between layers compressed) 
with respect to the expected wire grid at the readout end of the
BOS~98 prototype chamber. The uncompensated cross plate
deformations during assembly of each layer are indicated.
\label{res1}
}
\end{figure}

\begin{figure}[h]
\vspace{-8mm}

\hspace{-2mm}
%\mbox{\epsfig{figure=/home/iwsatlas1/kroha/tex/atlas/fortran/bos_2000_08_p2_r0_25_6p_layer_sct.eps,
%width=1.21\linewidth}}
\includegraphics[width=1.02\linewidth]{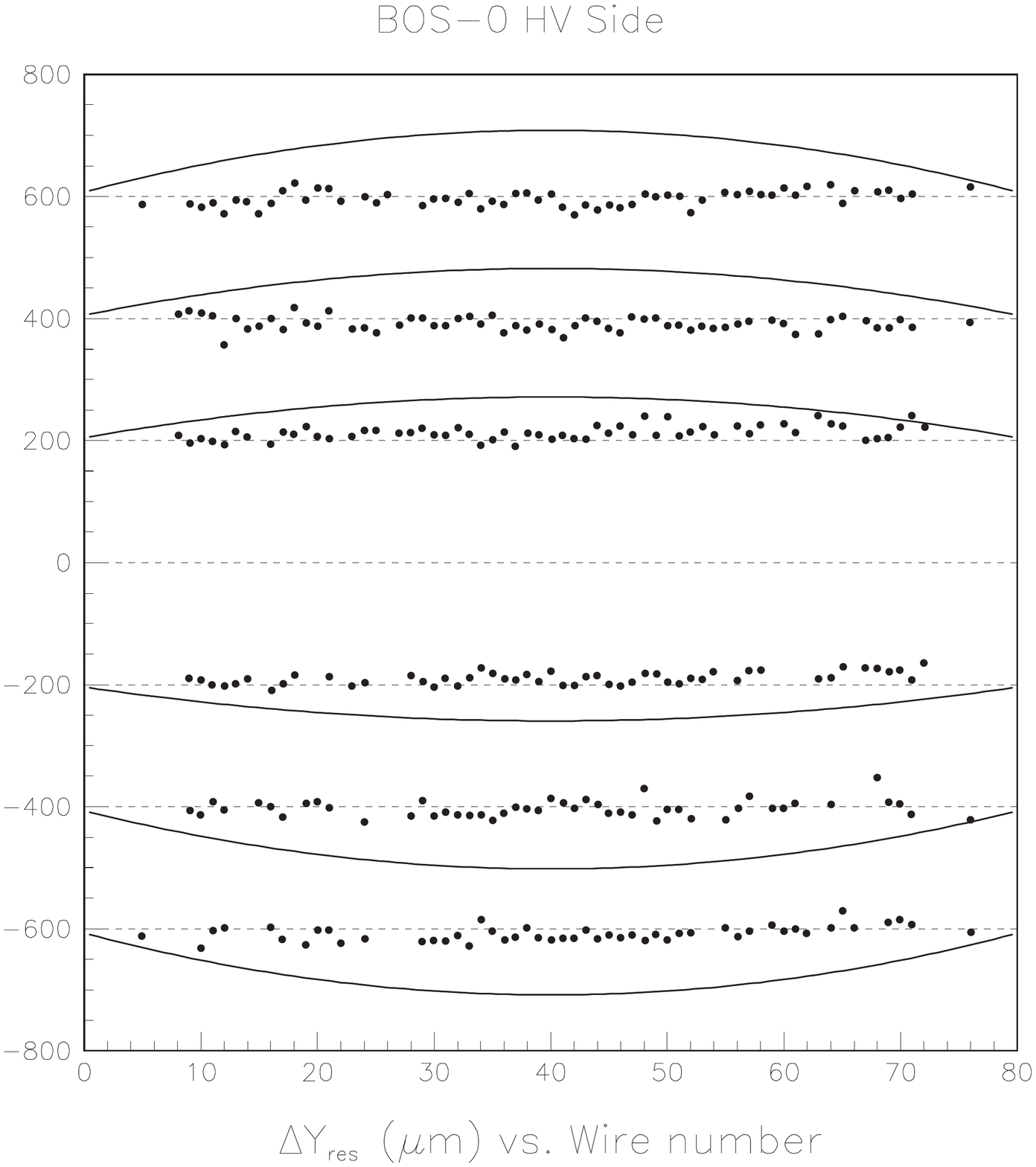}
\vspace{-15mm}

\caption{As Fig.~11 for the high-voltage end of the BOS-0 chamber.
\label{res2}
}
\end{figure}
\begin{figure}[h]
\vspace{-15mm}

\hspace{-52mm}
\begin{center}
%\mbox{\epsfig{figure=/.at/wwwatlas/ftp/outgoing/mdt/BOS-3800/bos98_tb_resol99_bw3.eps,
%width=1.04\linewidth}}
\includegraphics[width=1.02\linewidth]{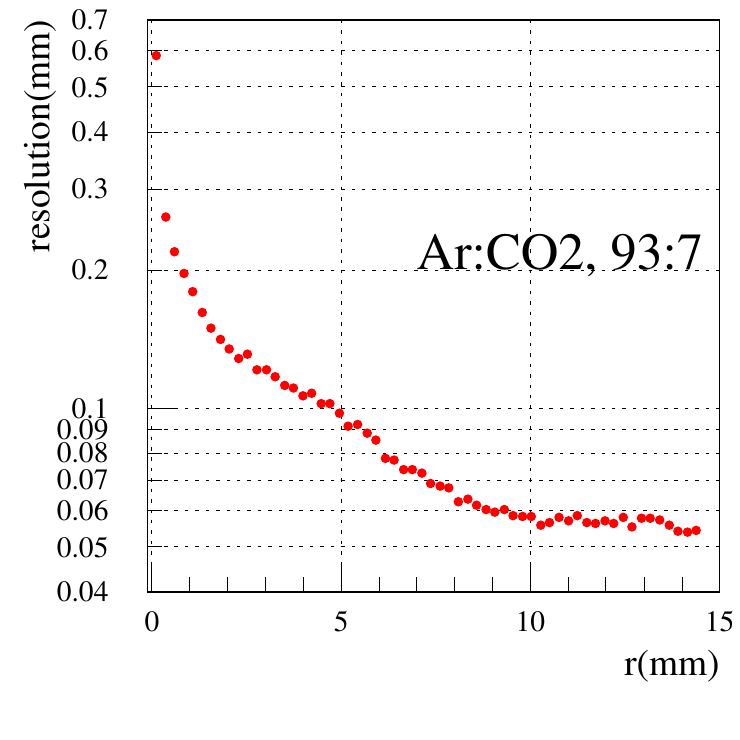}
\end{center}

\vspace{-10mm}
\caption{Single-tube resolution as a function of drift distance
for Ar:CO${}_2$ (93:7) gas mixture at 3 bar.\label{resol}}
\end{figure}

\begin{figure}[th]
\vspace{-30mm}

\begin{center}
%\mbox{\epsfig{figure=/.at/wwwatlas/ftp/outgoing/mdt/BOS-3800/bos98_tb_wirepos_bw2.eps,
%width=1.10\linewidth}}
\includegraphics[width=1.06\linewidth]{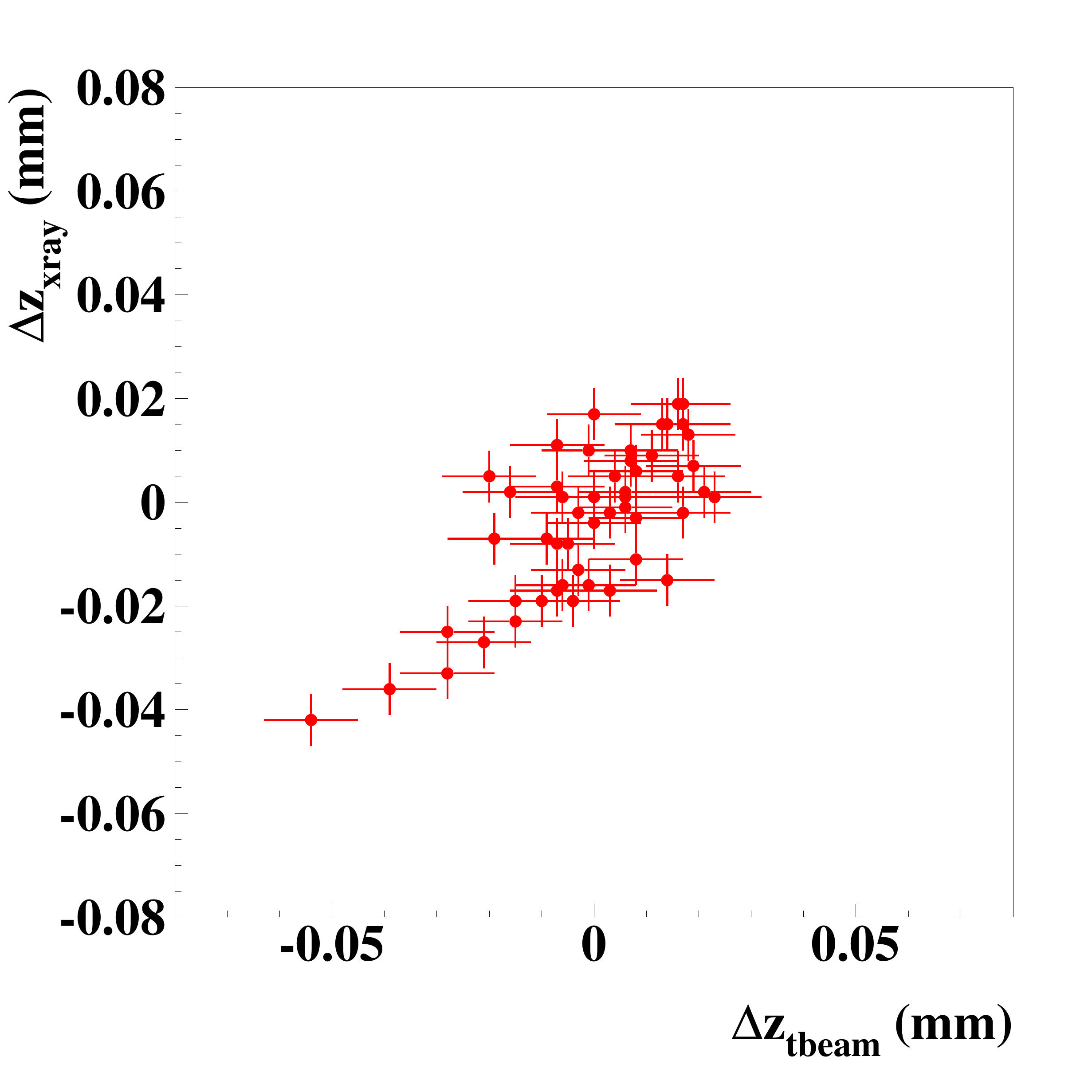}
\end{center}
\vspace{-5mm}

\caption{Correlation between the wire coordinate
measurements with X-rays and with
muon tracks from the test beam.\label{wirepos}}
\end{figure}

The MDT chambers (see Fig.~\ref{MDT}) consist of 3 or 4 layers of precise
aluminum drift tubes with $29.970\pm 0.015$~mm outer diameter and $400~\mu$m wall thickness
on either side of a space frame carrying an optical monitoring system to
correct for chamber deformations. The drift tubes are operated at a gas pressure of 3~bar 
to provide a single-tube position resolution of at least $80~\mu$m (rms)
with an Ar:CO$_2$ (93:7) gas mixture and at the low gas gain of $2\times 10^4$ required 
to prevent ageing of the drift tubes at the high background rates at the LHC.
The sense wires of the drift tubes have to be positioned
in individual tubes with an accuracy of $10~\mu$m (rms) and
in the whole chamber with an accuracy of $20~\mu$m (rms) in order to obtain a chamber position resolution of
$40~\mu$m (rms). 

In total 1200 MDT chambers containing 400000 drift tubes of 1--6~m length have to be constructed
for the ATLAS muon spectrometer at 13 production sites over a period of 4 years. In Munich, the
production of 88 of the largest MDT chambers with 432 drift tubes of 3.8~m length in 6 layers and 
with a width of 2.16~m has started. 
The drift tubes for these chambers are fabricated in a joint facility
at the Joint Institute for Nuclear Research (JINR) in Dubna, Russia.
The first MDT chamber for the ATLAS detector (named `BOS-0') has been completed in August 2000.

In spring 1998, the full-scale prototype of a MDT chamber of this type (named `BOS~98')
has been built with the methods developed for large-scale production~\cite{TDR}.
With the prototype chamber, it has first been demonstrated
that the required high mechanical accuracy can be achieved. Over the last two years, operation
experience with MDT chambers was gained with the prototype in the muon test beam at CERN.

\section{Drift Tube Fabrication}
For the prototype chamber, the sense wires were positioned and fixed at the tube ends
using external references for tubes and wire and fast-curing glue (see Fig.~\ref{glue}).
The effect of glue shrinkage on the wire position was measured and taken into account.
The endplug of the drift tubes designed for this method (see Fig.~\ref{endplug1}) 
does not require high precision in the fabrication.
After assembly, the wire positioning accuracy at the ends of the drift tubes with respect to the 
outer tube diameter was measured 
to be $10~\mu$m (rms) in both coordinates including the non-roundness of the tubes
(see Fig.~\ref{wp1_2d}) using a stereo X-ray technique with a resolution of $2~\mu$m.

For the large-scale production, a precisely machined endplug variant (see Fig.~\ref{endplug2})
has been adopted where the wire is located in a spiral hole
concentric with an aluminum reference ring
on which the drift tube ends are positioned during chamber assembly
(see below). The injection moulding process of the insulating plastic (Noryl)
body of the endplugs with the metal inserts has been carefully optimised
in order to prevent stresses and the development of cracks which can make the drift tubes leak.
The X-ray measurements show a wire positioning accuracy of $7~\mu$m (rms) (see Fig.~\ref{wp2_1d}).

Reliable ground contact of the aluminum tubes (the cathodes of the
drift tubes) is provided by spot welds to the aluminum ring on the endplug 
using a specially developed laser welding technique employing filler wire.
The contact resistance at a current of 10~mA stays below 1~m$\Omega$ even after
an accelerated corrosion test with exposure to salt spray for 48 hours.

The drift tubes for serial production are assembled semi-automatically in a clean room of class 10000
with temperature and relative humidity controlled to be $(20\pm 0.5)^\circ$C and $(50\pm 10)\%$, 
respectively. They have to fulfil stringent quality criteria which include wire position
measurement with X-rays within $\pm 2.5\sigma$ ($\sigma = 10~\mu$m) in both coordinates,
gas leak rate at 3~bar below $10^{-8}$~bar$\;\cdot\;$l/s,
wire tension within $\pm 5\%$ of the nominal value of 350~g, and high voltage leakage current
for Ar:CO$_2$ (93:7) at 3~bar and 3.4~kV below 8~nA. 
At present, the drift tube rejection rate is about $3\%$.

In the drift tubes of the BOS~98 and the BOS-0 chamber,
the wire tension, determined from the measurement of the oscillation frequency $\nu$ of the
$50~\mu$m diameter gold-plated tungsten-rhenium (97:3) wire,
is uniform within $2\%$ (rms). 

\section{Chamber Assembly}

During the assembly of a MDT chamber~\cite{TDR},
the ends of the drift tubes for each layer are positioned 
on precision aluminum combs with an accuracy of $3~\mu$m (rms) 
in horizontal ($z$) and vertical ($y$) direction.
The combs have been produced industrially by spark erosion
and are installed on a flat granite table in a clean room of class 100000
with temperature and relative humidity controlled to be $(20\pm 0.5)^\circ$C and $(50\pm 10)\%$, respectively.
Over their whole length the tubes are held straight in 9 parallel rows of combs with vacuum suction.

The tube layers inserted in the combs are glued subsequently to the aluminum space frame (three
cross plates connected by two long beams; see Fig.~\ref{table1}) which for this purpose
is positioned with respect to the combs with an accuracy of $\pm 5~\mu$m in $y$ and $z$
on precision towers at the ends of the three cross plates (see Fig.~\ref{table1}).
The positioning of the space frame
is monitored with laser beams and transparent optical position sensors~\cite{ALMY1}-\cite{ALMY4}.

While the space frame is supported on the
reference towers during glueing of a tube layer, the 2.16~m wide cross plates of the large chambers bend between
the support points under the weight of the chamber by up to $80~\mu$m at the ends and $100~\mu$m in the middle.
The gravitational sag of the cross plates during assembly 
is measured with optical sensors installed on the cross plates
(see Fig.~\ref{sag2}). In order to prevent
deformations of the tube layers after glueing them to the space frame, the cross plates are
also supported via the long beams applying forces with computer controlled
pneumatic actuators until the sag is compensated 
without lifting the chamber from the reference towers. 
After positioning the chamber on the reference towers,
the pneumatic actuators instantaneously apply the required forces
at the ends of 4 bars inserted through holes in the long beams close to the cross plates
as shown in Fig.~\ref{table1}.

\section{X-Ray Measurements of the Chambers}

The wire positions in the completed chambers have been measured at CERN~\cite{Xtomo}
with scans with stereo X-ray sources perpendicular to the wires.
The reproducibility of the X-ray measurements of the wire coordinates was within $5~\mu$m (rms)
during the scans of the prototype chamber in 1998 and
now is within $3~\mu$m (rms).

A fit of an ideal wire grid to the measured wire coordinates $y$ (perpendicular to the
tube layers) and $z$ (parallel to the tube layers) allows a determination of the geometrical
parameters of the chamber and to evaluate the wire positioning accuracy. 
In Table~\ref{fit} the fitted parameters, the horizontal ($z$) and vertical ($y$) wire pitch (at $20\pm 0.5^\circ$C)
and the $y$- and $z$-separations $\Delta y$ and $\Delta z$
between the two triple-layers, from the scans at the chamber ends are compared
to the design values for the prototype chamber (BOS~98) and for the first production chamber (BOS-0).
For the BOS~98 chamber, the X-ray results are the average of several scans at the same position along
the tubes. The $\Delta y$ values at both ends of this chamber differ because
of a known common $y$-offset of the wire positions in the tubes at the readout end.
During the assembly of the BOS-0 chamber, an unexpectedly large shrinkage of the glue 
between tube layers and frame was observed which is taken into account in the
design values. With the different type of glue used for the BOS~98 chamber,
no significant glue shrinkage was observed.
For both chambers, the fitted parameters agree very well with the design values.

Distributions of the residuals of the measured wire coordinates $y$ and $z$
with respect to the fitted grid 
are shown in Figs.~\ref{Xray1} and \ref{Xray2} for the two chambers.
The widths of the distributions for X-ray measurements at both ends of the chambers as well
as near the center are summarised in Table~\ref{width}.
The wire locations at intermediate
positions between the tube ends are determined by the wire locations at the ends (weighted averages)
and by the gravitational sag $s$ of the wires. The wire sag is known
from the measurement of the wire oscillation frequency $\nu$ via the relation $s ={g\over 32\nu^2}$ where $g$
is the gravitational acceleration. Variations in the maximum wire sag of $2\%$ (rms)
of the nominal $195~\mu$m due to the variations of the wire tension are negligible.

The residuals of the $y$-coordinates are shown in Figs.~\ref{res1} and \ref{res2}.
The gravitational sag of the cross plates which has to be compensated for the assembly
of each tube layer (see Fig.~\ref{sag2}) is indicated.
The BOS-0 chamber, for cost reasons, has less stiff cross plates than the prototype chamber
and therefore larger cross plate sag. The statistical fluctuations of the wire locations
are larger in the prototype chamber because the aluminum tube walls,
instead of the precise endplugs, were used as references
for the positioning of the drift tubes 

The wire positioning accuracy in the prototype chamber is better than the required $20~\mu$m (rms).
With the first production chamber, a wire positioning accuracy of better than $14~\mu$m (rms)
has been achieved for one of the largest chamber types in the ATLAS muon spectrometer.
The main improvement with respect to wire positioning accuracy compared to the 
prototype chamber is the introduction of the aluminum reference ring on the
endplugs which allows more precise positioning of the drift tube ends on the assembly combs
than the aluminum tube walls used before.
Based on the present experience, a wire positioning accuracy of about $10~\mu$m (rms) 
is reachable for such large chambers.

\section{Test Beam Measurements}

The prototype chamber has been tested in a 300~GeV muon beam at CERN 
at perpendicular incidence to the tube layers. 
Ar:CO${}_2$ $(93:7)$ at 3~bar was used as drift gas.
Using a silicon strip detector telescope as external reference, 
the space to drift-time relationship and the
position resolution as a function of the drift distance r have been determined.
(see Fig.~\ref{resol}). The average single-tube resolution at the low interaction rates 
is $70~\mu$m (rms).

The r-t-relationship measured locally
was applied to the other drift tubes in the beam allowing only the
maximum drift time to vary within $\pm 6$~ns (rms) because of varying operating conditions.
Requiring the track residual distributions as function of the drift distance 
to be symmetric left and right of the wires
provides information about displacements of the wires from their nominal positions
in $z$-direction with respect to a reference wire.
Comparison with the X-ray measurements of the $z$-coordinates 
of the wires shows a good correlation (see Fig.~\ref{wirepos}). 
Both measurements agree within $10~\mu$m (rms).

\section{Acknowledgements}

We wish to thank our collegues at the JINR, Dubna for the fabrication of such excellent drift tubes
for the BOS-0 chamber and are indebted to the X-ray tomograph group at CERN
for the prompt and accurate measurements of our chambers.

\end{document}